\documentclass[10pt]{article}
\usepackage{opex3}
\usepackage{color}

\usepackage{upgreek}

\usepackage{siunitx}
\DeclareSIUnit\gauss{G}

\begin{document}
\title{Calibrating High Intensity Absorption Imaging of Ultracold Atoms}

\author{Klaus Hueck,$^{1,*}$ Niclas Luick,$^1$ Lennart Sobirey,$^1$ Jonas Siegl,$^1$ Thomas Lompe$^1$ and Henning Moritz$^1$\newline Logan W. Clark$^2$ and Cheng Chin$^2$}

\address{$^1$Institut f\"ur Laserphysik, University of Hamburg, Luruper Chaussee 149, 22761, Hamburg, Germany\newline$^2$James Franck Institute, Enrico Fermi Institute and Department of Physics, University of Chicago, 929 E 57th Street GCIS ESB11, Chicago, IL 60637, USA}

\email{$^*$klaus.hueck@physik.uni-hamburg.de} 



\begin{abstract}
Absorption imaging of ultracold atoms is the foundation for quantitative extraction of information from experiments with ultracold atoms.
Due to the limited exposure time available in these systems, the signal-to-noise ratio is largest for high intensity absorption imaging where the intensity of the imaging light is on the order of the saturation intensity.
In this case, the absolute value of the intensity of the imaging light enters as an additional parameter making it more sensitive to systematic errors.
Here, we present a novel and robust technique to determine the imaging beam intensity in units of the effective saturation intensity to better than \SI{5}{\percent}.
We do this by measuring the momentum transferred to the atoms by the imaging light while varying its intensity.
We further utilize the method to quantify the purity of the polarization of the imaging light and to determine the correct imaging detuning.
\end{abstract}



\section{Introduction}

Ultracold atoms have added a completely new toolset to the study of quantum many-body systems by allowing direct imaging of density and momentum distributions. 
This led to the observation of striking features such as the bimodal momentum distribution signifying the onset of Bose-Einstein condensation \cite{Anderson95,Davis95} or vortex lattices in rotating superfluids\cite{Matthews99,Abo-Shaeer01}. 
Ultracold atoms have also been used for precision measurements such as the extraction of the equation of state \cite{Hung11,Ku12} from an accurate measurement of the density distribution of the system.
The most common method to measure such density distributions is absorption imaging: 
the atoms are illuminated with a short pulse of resonant light and the shadow cast by the atoms is imaged onto a camera.
The fraction of light transmitted through the system decreases exponentially according to the Beer-Lambert law, provided the atomic transition is not saturated.
In this case the density distribution integrated along the imaging direction $n_{2D}$ can be extracted from a relative measurement which compares images with and without atoms.
Hence, no knowledge of the absolute intensities is required. 

However, absorption imaging with intensities well below the saturation intensity has clear limitations. 
One prominent issue is the imaging of optically dense clouds, where the number of photons transmitted through the sample and hence the signal-to-noise ratio becomes very small. 
This can either be solved by using techniques such as phase contrast imaging \cite{Stamper-Kurn99} or by simply saturating the transition so that more photons are transmitted\cite{Reinaudi07,Hung11,Yefsah11}. 
Imaging in the saturated regime is also helpful when a high spatial resolution is desired. 
To this end, it is essential to minimize the motion of atoms caused by the recoil from scattered photons during the imaging pulse. 
The motional blurring is minimized by utilizing short imaging times with high intensities.

For absorption imaging with higher intensities a new scale appears: the saturation intensity $I_{\textit{sat}}$. 
This implies that the column density can no longer be determined from a purely relative measurement of two intensities. 
Rather, the absolute intensities transmitted with and without atoms present have to be known in order to extract an atomic density.
However, determining the imaging beam intensity $I$ at the position of the atoms is prone to systematic errors. 
Additionally, imperfect polarization of the imaging light can lead to a reduced scattering cross-section and hence an increased effective saturation intensity $I_{\textit{sat}}^{\textit{eff}}$.

Here, we present a novel technique which allows for a precise and robust calibration of high intensity absorption imaging.
It is based on measuring the momentum transferred to the atoms by the imaging pulse which is directly proportional to the number of scattered photons.
The beam intensity $I$ for which the transferred momentum and hence the photon scattering rate is equal to half its maximum value corresponds to the effective saturation intensity $I_{\textit{sat}}^{\textit{eff}}$.
Measuring the momentum transfer also allows to determine the polarization purity of the imaging light at the position of the atoms as well as the detuning $\Delta$.

The manuscript is organized as follows: 
In section two a brief review of strong saturation imaging is given and section three describes the experimental setup. 
In the fourth section, we describe our method to measure the momentum transfer and apply it to determine $I/I_{\textit{sat}}^{\textit{eff}}$, the correct detuning and the polarization purity of the imaging beam.
In section five we summarize the results. 

\section{High intensity imaging}
Here, we provide a short summary of the theory of resonant saturated absorption imaging. 
When the saturation of an optical transition becomes relevant, the Beer-Lambert law is modified to 
\begin{equation}
  \frac{dI(x,y,z)}{dz} = -n(x,y,z)\sigma_{\textit{eff}}\frac{1}{1+I(x,y,z)/I_{\textit{sat}}^{\textit{eff}}}I(x,y,z), 
  \label{eqn:ModLambertBeer}
\end{equation}
where $n$ is the atom density and $I(x,y,z)$ the intensity at position $(x,y)$ of the imaging light propagating along the $z$-direction \cite{Reinaudi07}.
$\sigma_{\textit{eff}}=\frac{\sigma_0}{\alpha}$ and $I_{\textit{sat}}^{\textit{eff}}=\alpha I_{\textit{sat}}$ are the effective cross-section and effective saturation intensity. 
Here, a parameter $\alpha > 1$ was introduced to capture effects of non-perfect polarization or magnetic field orientation which reduces the cross-section and saturation intensity in a two-level system from its ideal values $\sigma_0 = \frac{3\lambda^2}{2\pi}$ and $I_{\textit{sat}}^0 = \frac{\pi}{3}\frac{hc\Gamma}{\lambda^3}$\cite{Foot05}. 
Here, $h$ is Planck's constant, $c$ the speed of light, $\lambda$ is the wavelength and $\Gamma$ the natural linewidth of the imaging transition.
Integration of Eq.~(\ref{eqn:ModLambertBeer}) along $z$ then yields the optical column density
\begin{equation}
  od(x,y) = \sigma_{\textit{eff}}n_{2D}(x,y) = \sigma_{\textit{eff}}\int_{-\infty}^{\infty}n(x,y,z)dz = -\underbrace{\ln\left(\frac{I_{\textit{out}}(x,y)}{I_{\textit{in}}(x,y)}\right)}_{\text{Log-Term}}+\underbrace{\frac{I_{\textit{in}}(x,y)-I_{\textit{out}}(x,y)}{I_{\textit{sat}}^{\textit{eff}}}}_{\text{Lin-Term}},
  \label{eqn:2DDensity}
\end{equation}
with $I_{\textit{in}}$ and $I_{\textit{out}}$ being incident and transmitted intensities, respectively. 
When the intensities are small compared to the saturation intensity the linear term can be neglected and we recover the simple Beer-Lambert law which only depends on relative intensities. 
However, for higher intensities the linear term becomes significant and the intensities have to be known in units of the effective saturation intensity.

Furthermore, the intensities $I_{\textit{in}}(x,y)$ and $I_{\textit{out}}(x,y)$ read out from a pixel $(i,j)$ of the camera are given in units of count rates, which will be denoted $C_{\textit{in}}(i,j)$ and $C_{\textit{out}}(i,j)$ in the following. 
The count rate is related to the intensity by 
\begin{equation}
  C=\frac{IA_{\textit{pix}}/M^2}{hc/\lambda}\times T\times QE\times G,
  \label{eqn:CountToIntensity}
\end{equation}
where $A_{\textit{pix}}$ is the area of a camera pixel, $M$ the magnification of the imaging system, $T$ the transmission through the imaging system, $QE$ the quantum efficiency and $G$ the conversion factor between counts and photo-electrons of the camera. 
Expressing Eq.~(\ref{eqn:2DDensity}) in counts $C$ leads to 
\begin{equation}
  \sigma_{\textit{eff}}n_{2D}(i,j) = -\ln\left(\frac{C_{\textit{out}}(i,j)}{C_{\textit{in}}(i,j)}\right)+\frac{C_{\textit{in}}(i,j)-C_{\textit{out}}(i,j)}{C_{\textit{sat}}^{\textit{eff}}}.
  \label{eqn:2DDensityC}
\end{equation}
It becomes apparent that with an independent measurement of $C_{\textit{sat}}^{\textit{eff}}$ there is no need to know $T$, $QE$ and $G$.
The problem hence reduces to finding the number of counts $C_{\textit{sat}}^{\textit{eff}}$ on the camera corresponding to the effective saturation intensity and the effective cross-section $\sigma_{\textit{eff}}$.
Knowledge of $\sigma_{\textit{eff}}$ is also necessary for low intensity imaging.

\begin{figure}
  \center
  \includegraphics[width=.9\linewidth]{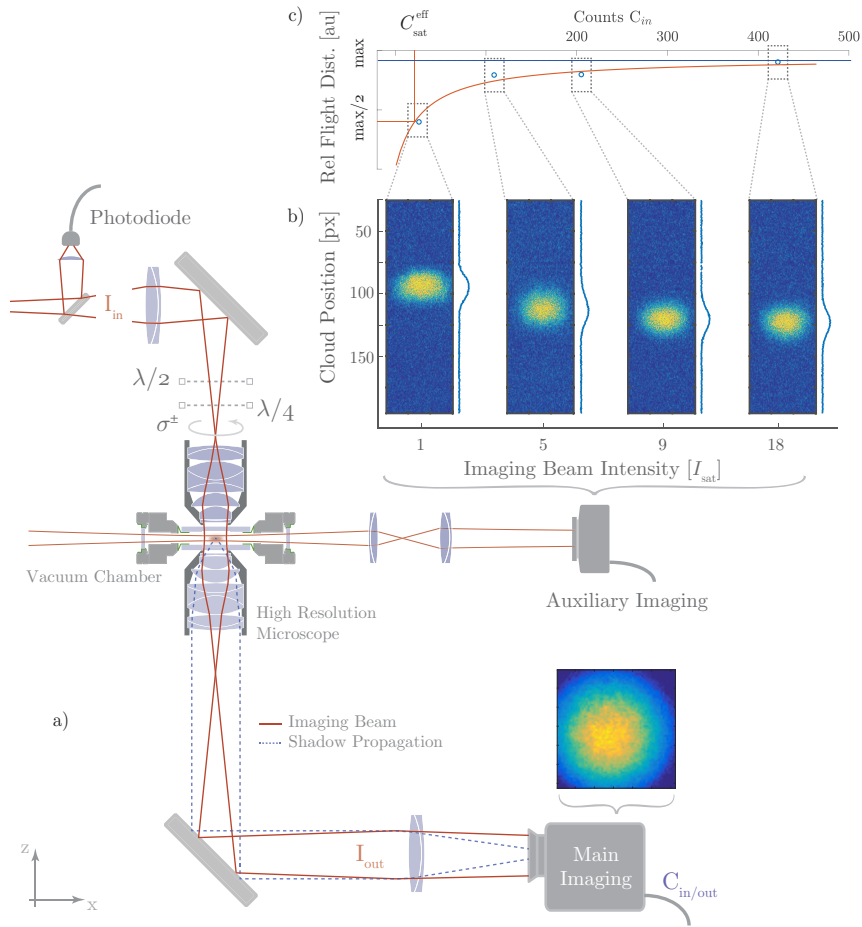}
  \caption{\label{fig:ExpSetup} Sketch of the experimental setup: a) Absorption imaging of a cloud of ultracold atoms trapped inside a vacuum chamber. Imaging of the cloud is possible along two axes, the main axis (z-axis) and an auxiliary imaging axis (x-axis). By flashing on the main imaging beam, the atoms accelerate in z-direction. After some time of flight their position is recorded with the auxiliary imaging. b) Density distributions after \SI{80}{\micro\second} time of flight imaged along the auxiliary imaging direction. The flight distances shown in c) are extracted from the images and plotted as a function of the intensity of the imaging pulse, which is measured on the main imaging camera. From the saturation of the flight distance we can determine the count rate on the main imaging camera which corresponds to the effective saturation intensity.}
\end{figure}

\section{Experimental setup}
The experimental setup is designed to produce, manipulate and image two-dimensional superfluids of $^6$Li atoms with high spatial resolution\cite{Weimer15}.
The imaging system along the $z$-direction has a numerical aperture of 0.62 and thus a very limited depth of field of about \SI{2}{\micro\meter}. 
Hence, short imaging pulses are required to minimize the motion along the optical axis as well as the random walk in $xy$-plane during the imaging process in order to avoid motion blurring.
In our $^6$Li experiment we typically use \SI{5}{\micro\second} imaging pulses with $s_0 = 1$.
This causes motional blurring of about \SI{1.5}{\micro\meter}\cite{Stamper-Kurn99}.
Under this condition it is therefore advantageous to use high intensity imaging to maximize the number of scattered photons, since the signal-to-noise ratio in absorption imaging is typically limited by photon shot noise.
The quantization axis for the atoms is given by a magnetic field parallel to the optical axis.
Therefore, the absorption cross-section on the $^2$S$_{1/2}$, F=1/2 to $^2$P$_{3/2}$, F=3/2 imaging transition is maximal for circularly polarized light.

The imaging laser light is generated by a frequency stabilized narrow linewidth diode laser which is frequency shifted by an acousto-optical deflector (AOD) in double pass configuration.
This AOD setup enables chirping the imaging frequency with high ramp rates exceeding \SI{5}{\mega\hertz\per\micro\second} using a direct digital synthesis (DDS) frequency source.
The imaging light intensity is controlled via a sample and hold feedback loop. 

For the measurements presented in the following, we work with about \num{60e3} atoms per spin state in the two lowest hyper fine states of $^6$Li at a magnetic offset field of \SI{950}{\gauss}. 
The atoms are magnetically trapped in the radial direction (trapping frequencies $\nu_x\approx\nu_y \approx \SI{27}{\hertz}$). 
The tight confinement in $z$-direction is realized by a highly elliptic dipole trap featuring a trapping frequency of about \SI{0.5}{\kilo\hertz}.
In order to avoid multiple scattering effects during the calibration measurements the optical density $od(x,y)$ along the $z$-direction as well as the atomic density $n(x,y,z)$ is chosen to be low.
We note that optically thin samples $od(x,y)<1$ are only required for calibration.
Once the effective saturation count rate $C_{\textit{sat}}^{\textit{eff}}$ is known, optically thick samples can be imaged by using high intensities to saturate the imaging transition.
However, to avoid collective scattering effects during the imaging, the condition that the inter-particle spacing $1/n(x,y,z)$ has to be larger than the imaging wavelength $\lambda$ still has to be fulfilled\cite{Chomaz12}.

\section{Method}
\subsection{Previous work}
There exist two closely related approaches to determine $I_{\textit{sat}}^{\textit{eff}}$ or $C_{\textit{sat}}^{\textit{eff}}$ in the literature \cite{Ries15,Reinaudi07,Horikoshi16}, which rely on taking absorption images of clouds with constant atom numbers while scanning over a wide range of imaging intensities $I_{\textit{in}}$. 
For an assumed value of $\alpha = \frac{\sigma_0}{\sigma^{\textit{eff}}}$ the total atom number $N(I_{\textit{in}},\alpha)$ for each intensity can be determined by evaluating Eq.~(\ref{eqn:2DDensity}) or (\ref{eqn:2DDensityC}) for all pixels and integrating over the atomic density distribution. 

Ries et al. \cite{Ries15} and Horikoshi et al. \cite{Horikoshi16} determine $C_{\textit{sat}}^{\textit{eff}}$ by requiring $N(I_{\textit{in}},\alpha)$ to be independent of $I_{\textit{in}}$.
$\alpha$ is still a free parameter and has to be determined by an independent measurement, e.g. by comparing the measured equation of state to a theory prediction \cite{Horikoshi16}.

Reinaudi et al. \cite{Reinaudi07} calculate the intensity of the imaging beam $I_{\textit{in}}$ at the position of the atoms from measurements of the imaging beam power and the transmission of the imaging system. 
The parameter $\alpha$ is found by requiring $N(I_{\textit{in}},\alpha)$ to be independent of $I_{\textit{in}}$. 

\subsection{This work}
The method described in this work neither requires a power measurement of the imaging beam nor a constant atom number during the calibration process.
The basic principle is depicted in Fig.~\ref{fig:ExpSetup}. 
We accelerate the cloud of atoms in $z$-direction by illuminating it with a short pulse from the main imaging beam after the dipole traps are switched off.
The momentum transferred to each atom during the pulse is proportional to the number of scattered photons.
Subsequently, the transferred momentum is determined by recording the flight distance of the accelerated cloud after some time of flight with an auxiliary imaging system along an orthogonal direction.
The momentum transferred to the atoms saturates with increasing intensity $I_{\textit{in}}$ since the resonant photon scattering rate $\gamma$ saturates according to
\begin{equation}
  \gamma(s) = \frac{\Gamma}{2}\frac{s_0}{1+s_0}. 
  \label{eqn:ResScatRate}
\end{equation}
Here, $\Gamma$ is the natural linewidth of the imaging transition and $s_0=\frac{I_{\textit{in}}}{I_{\textit{sat}}^{\textit{eff}}} = \frac{C_{\textit{in}}}{C_{\textit{sat}}^{\textit{eff}}}$ denotes the on resonant saturation parameter which is given by the ratio of the imaging beam intensity to the effective saturation intensity.
The intensity for which the atoms traveled half their maximum flight distance corresponds to $I_{\textit{sat}}^{\textit{eff}}$. 
Hence, the magnification of the auxiliary imaging system drops out of the determination of the effective saturation intensity and therefore does not need to be known.
Note that Eq.~(\ref{eqn:ResScatRate}) only holds true if the imaging beam is resonant with the atoms.

\subsection{Detuning determination}
To determine the resonance frequency we apply imaging pulses with different detunings and maximize the transferred momentum to the atoms.
\begin{figure}
  \center
  \includegraphics[width=\linewidth]{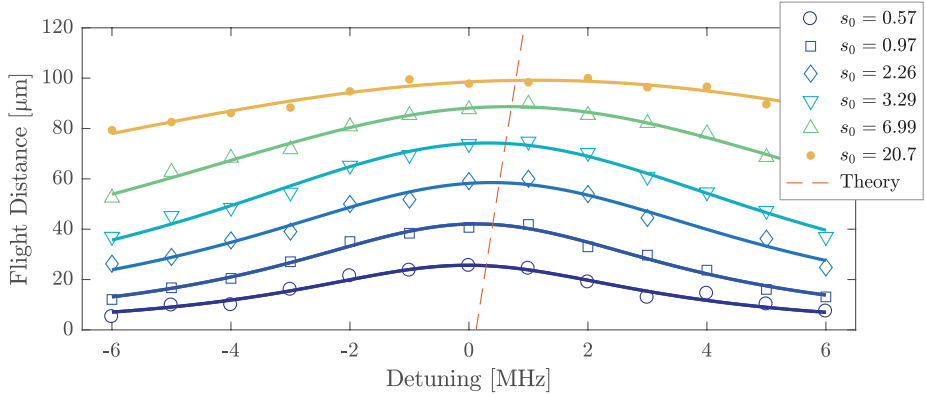}
  \caption{\label{fig:Detuning} Flight distance vs. laser detuning: 
  The resonance frequency of the imaging laser is found by varying the detuning and thereby maximizing the flight distance of the atoms after illumination with a \SI{1}{\micro\second} long imaging pulse. 
  A Lorentzian fit (solid lines) to the data gives the center frequency as well as the width of the imaging transition, which shows power broadening for high intensities.
  The saturation parameter $s_0=I/I_{\textit{sat}}^{\textit{eff}}$ for the different curves is extracted from the fits.
  The red dashed line indicates the theoretically expected position of the maxima which are affected by the accumulated Doppler shift during the imaging pulse.
  The x-axis is offset by the fitted resonance frequency.}
\end{figure}
A \SI{1}{\micro\second} long pulse of imaging light along the $z$-direction is applied to accelerate the atom cloud and after \SI{80}{\micro\second} time of flight, the position of the atom cloud is recorded using the auxiliary imaging. 
Then, the center of mass of the atom cloud is found by performing a Gaussian fit to the recorded density profile. 
The resulting center positions are plotted in Fig.~\ref{fig:Detuning} for different detunings and imaging intensities. 
The functional form is expected to be captured by a Lorentzian curve resembling the resonance behavior of the atomic transition. 
A fit of the form 
\begin{equation}
  z(\nu_L, s) = z_0+\eta\frac{\Gamma}{2}\frac{s_0}{1+s_0+\left(\frac{2\Delta}{\Gamma}\right)^2} 
  \label{eqn:ScatRateArb}
\end{equation}
yields the detuning $\Delta = \nu_L-\nu_A$ between the atomic resonance frequency $\nu_A$ and the laser frequency $\nu_L$. Here, $\eta$ is the conversion from position difference on the camera to the scattering rate and is left as a fit parameter as well as the saturation parameter $s_0$, and $z_0$ is the initial position of the atoms and $\Gamma$ the linewidth of the imaging transition.
As shown in Fig.~\ref{fig:Detuning} the results are well described by Eq.~(\ref{fig:Detuning}).
For higher saturation parameters $s_0$ the resonance is blue-shifted since the atoms accumulate a Doppler shift during the acceleration (see section~\ref{sec:Doppler}).
However, for imaging intensities $s_0\le1$ and illumination times $t\le1\, \si{\micro\second}$ the shift of the fitted detuning is smaller than the fitting error.
The fitting error of the detuning is typically below \SI{200}{\kilo\hertz} which is less than \SI{4}{\percent} of the linewidth of the $^6$Li D2 line. 

Our method has two clear advantages over the common method of finding the resonant laser frequency by maximizing the apparent atom number as determined by Eq.~(\ref{eqn:2DDensity}).
It is independent of fluctuations in the atoms number.
Furthermore, it is not influenced by effects such as lensing in a dense atom cloud imaged off-resonance which can systematically shift the apparent atom number maximum away from the correct resonance frequency.

\begin{figure}
  \center
  \includegraphics[scale=1]{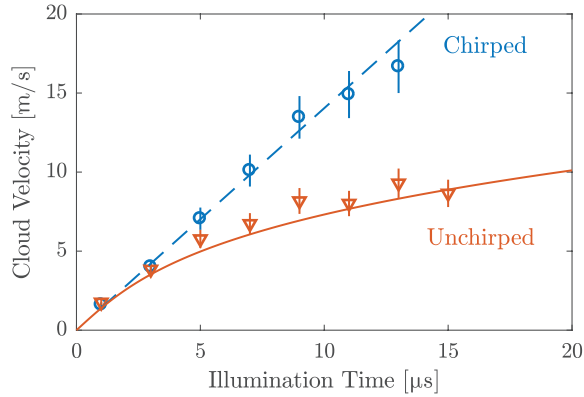}
  \caption{\label{fig:ChirpEff} Chirp efficiency: 
  Velocity of the atom cloud after accelerating it with the $z$-imaging beam with different illumination times at $I_{\textit{in}}=3.75I_{\textit{sat}}^{\textit{eff}}$.
  The effect of chirping the imaging beam frequency to compensate for the Doppler shift is clearly visible. 
  In the unchirped case (red triangles), the scattering rate decreases with illumination time and therefore the increase in cloud velocity becomes nonlinear. 
  In the chirped case (blue open circles) acceleration is constant.
  The lines indicate the theoretically expected cloud velocity for the unchirped (red solid) and chirped case (blue dashed). 
  The error bars indicate the systematic error estimated for the magnification of the auxiliary imaging.}
\end{figure}

\subsection{Doppler shift compensation}
\label{sec:Doppler}
For $^6$Li the small mass causes the atoms to acquire a significant Doppler shift of $\Gamma/2$ in only \SI{2.5}{\micro\second} when imaging with saturation intensity.
To be able to scatter more photons and hence increase the signal-to-noise ratio we compensate for the Doppler shift by performing a frequency chirp on the imaging laser frequency during the imaging. 
To find the chirp rate $m$ which exactly compensates for the Doppler shift we adjust the chirp rate such that we maximize the momentum transferred by the imaging pulse to the atoms. 
A typical chirp rate required for $^6$Li is on the order of $\SI{1.5}{\mega\hertz\per\micro\second}$ for $I/I_{\textit{sat}}^{\textit{eff}} = 1$.

We verify the found chirp rate by accelerating the cloud of atoms with an imaging beam pulse with varying lengths up to \SI{15}{\micro\second} and an intensity of $I_{\textit{in}} = 3.75I_{\textit{sat}}^{\textit{eff}}$. 
The velocity of the cloud after the acceleration is determined from position measurements after two different times of flight.
The experimental result as well as a theory calculation is shown in Fig.~\ref{fig:ChirpEff}.
Without chirping the laser frequency the atoms are Doppler shifted out of resonance during the illumination, the scattering rate decreases and the increase in velocity becomes nonlinear (red triangles). 
In the chirped case (blue circles) this effect vanishes and the increase in velocity scales linearly with the illumination time, which indicates a constant photon scattering rate.
In both cases the accumulated velocity is in excellent agreement with the theoretical expectation (dashed and solid lines). 

\begin{figure}
  \center
  \includegraphics[width=.9\linewidth]{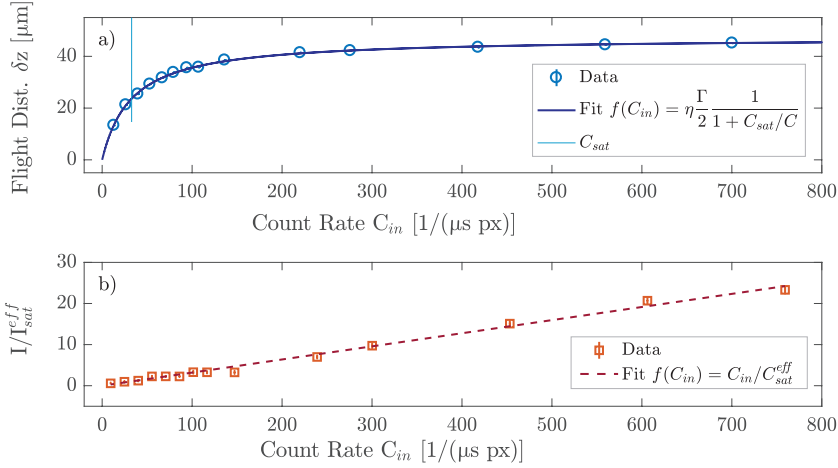}
  \caption{\label{fig:Isat} Determination of the effective saturation count rate $C_{\textit{sat}}^{\textit{eff}}$: 
  a)
  The difference in position for two times of flight ($\Delta t = \SI{10}{\micro\second}$) is shown as a function of imaging beam intensity and hence $C_{\textit{in}}$. 
  The velocity of the cloud saturates for higher imaging beam intensities and the saturation counts are determined by fitting Eq.~(\ref{eqn:ResScatRate}) to the data. 
  The resulting effective saturation count rate is $C_{\textit{sat,k}}^{\textit{eff}}=(\num{32.7+-.6})\,\text{(px\ $\upmu$s)}^{-1}$. 
  b) 
  The saturation parameter $s_0$ extracted from a Lorentzian fit to the power broadened spectra presented in Fig.~\ref{fig:Detuning} is plotted as a function of the count rate $C_{\textit{in}}$ on the main camera.
  The error bars represent the Lorentzian fit error. 
  A linear fit to the data yields $C_{\textit{sat,s}}^{\textit{eff}}=(\num{31.5+-1.3})\,\text{(px\ $\upmu$s)}^{-1}$.}
\end{figure}

\subsection{Effective saturation count rate}
Here, we present two independent methods to determine the effective saturation count rate $C_{\textit{sat}}^{\textit{eff}}$.
For both methods we utilize the measurement of momentum transferred from the imaging beam to the atoms.  
The first method is based on the saturation of the photon scattering rate for higher imaging beam intensities, while the second relies on the power broadening of the optical transition. 
Both methods feature errors below \SI{5}{\percent} and are in agreement to each other. 
Furthermore, the first method has also been used successfully in an experiment using ultracold $^{133}$Cs\cite{Hung11}.

For the first method the position difference after two times of flight of the atom cloud is measured for a pulse duration of \SI{5}{\micro\second} and different imaging beam intensities (see Fig.~\ref{fig:Isat} upper panel).
Two different times of flight are used in order to determine the mean velocity of the cloud after the momentum transfer.
This ensures, that only relative times and flight distances are compared and potential timing offsets cancel out.
Fitting the measured flight distance using Eq.~(\ref{eqn:ScatRateArb}) with the detuning fixed to $\Delta = 0$ and $C_{\textit{sat,k}}^{\textit{eff}}$ and $\eta$ as free parameters yields the saturation count rate for the system to high precision.
In addition, the fitted value for $\eta$ can also be used to extract the magnification of the auxiliary imaging system. 

For the second method, $C_{\textit{sat}}^{\textit{eff}}$ is determined from the power broadening of the imaging transition. 
To do this we use the fitted saturation parameters $s_0$ from the detuning measurements shown in Fig.~\ref{fig:Detuning}.
We plot the values for $s_0$ against their corresponding counts $C_{\textit{in}}$ on the main camera in the lower panel of Fig.~\ref{fig:Isat}.
A linear fit to the values for $s_0$ directly results in $C_{\textit{sat,s}}^{\textit{eff}}$.
In the following we use the mean value $C_{\textit{sat}}^{\textit{eff}} = (C_{\textit{sat,s}}^{\textit{eff}} + C_{\textit{sat,k}}^{\textit{eff}})/2$ of the two measurements.

\subsection{Polarization purity}
Since we image on a $\sigma^-$-transition along the magnetic field axis, any deviation from perfectly circularly polarized light reduces the effective absorption cross-section.
To check the purity of the polarization we use the orthogonal polarization $\sigma^+$, for which $\sigma_{\textit{eff}}$ should be zero.
First, we adjust for $\sigma^+$-polarization by minimizing the residual acceleration at a high beam intensity of $I^+=25I_{\textit{sat}}^{\textit{eff}}$.
This residual acceleration is directly proportional to the number of scattered photons.
We then set the polarization back to $\sigma^-$ and find the value of the imaging beam intensity $I^-$ for which we scatter the same amount of photons.
The purity of the polarization is now simply given by the ratio of the two intensities $p = I^-/I^+$. 
This results in a lower bound for the polarization purity of $p>\SI{99.6}{\percent}$.
This includes a possible mismatch between the magnetic field axis and the propagation direction of the imaging light. 
Hence, it is valid to approximate $\alpha=1$.

\subsection{Validation}
\begin{figure}
  \center
  \includegraphics[width=.8\linewidth]{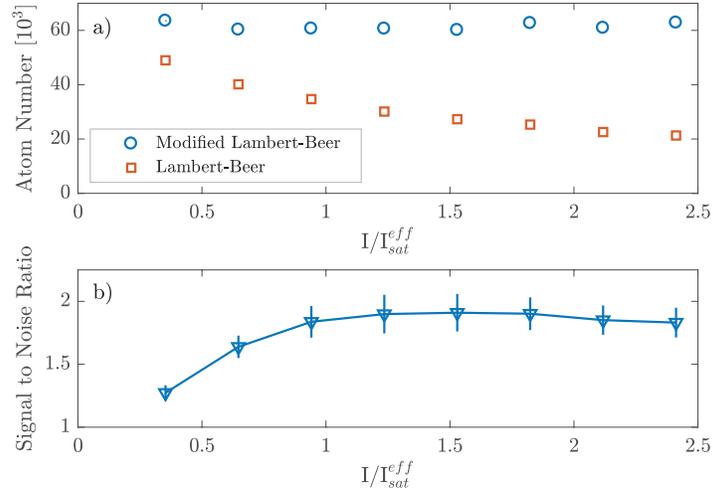}
  \caption{\label{fig:AtomRef} Proof of validity and optimization of signal-to-noise ratio: 
  a) 
  Extracted atom number as a function of imaging beam intensity using the modified (blue circles) and unmodified (red squares) Beer-Lambert law. 
  Each data-point is an average of about $\approx$ 70 measurements.
  The error-bars are smaller than the size of the symbols. 
  When evaluating the data with the modified Beer-Lambert law the atom number does not depend on the imaging intensity.
  This validates our method to determine the value of the effective saturation intensity $I_{\textit{sat}}^{\textit{eff}}$.
  For low imaging intensities the result from the unmodified Beer-Lambert law approaches the correct atom number.
  b) The signal-to-noise ratio evaluated on a single pixel basis is maximized for intensities of $I=1.5I_{\textit{sat}}^{\textit{eff}}$. The blue line is a guide to the eye.}
\end{figure}

Finally, we validate the effective saturation count rate $C_{\textit{sat}}^{\textit{eff}}$ against the experiment.
For that, we take density images of identically prepared clouds with different imaging beam intensities and extract atom numbers according to Eq.~(\ref{eqn:2DDensityC}) using the previously determined values of $C_{\textit{sat}}^{\textit{eff}}$ and $\alpha$.
If $C_{\textit{sat}}^{\textit{eff}}$ was determined correctly, Eq.~(\ref{eqn:2DDensityC}) should result in constant densities regardless the imaging beam intensity used.
We find the extracted atom number to be independent of $I_{\textit{in}}$ as can be seen in the upper panel in Fig.~\ref{fig:AtomRef}.
This validates our method of determining $C_{\textit{sat}}^{\textit{eff}}$. 

Furthermore, we evaluate the signal-to-noise ratio of the absorption images by evaluating the single pixel standard deviation for 70 subsequently taken images (Fig.~\ref{fig:AtomRef}, lower panel, blue triangles). 
The signal-to-noise ratio is maximized for intensities around $I_{\textit{in}}\approx 1.5I_{\textit{sat}}^{\textit{eff}}$.

\section{Conclusion}
In this work we have presented a novel technique to calibrate all parameters necessary to measure atomic densities using high intensity absorption imaging. 
These parameters include the detuning $\Delta$, the effective saturation count rate $C_{\textit{sat}}^{\textit{eff}}$, the polarization purity $p$ and hence $\alpha$ and finally the chirp rate required to compensate for Doppler shifts during the imaging.  
This allows for an intensity independent determination of the atomic column density $n_{2D}$. 
Thus, the intensity can freely be set to a value maximizing the signal-to-noise ratio.
The presented method only relies on relative measurements of the momentum transferred to the atoms by illuminating them with a resonant laser beam which makes it robust against systematic errors.
This makes the presented method a valuable tool for precision measurements of ultracold gases, especially as it is easily adapted to other experiments using different atomic species.

\section*{Funding}
The research leading to these results has received funding from the European Union's Seventh Framework Programme (FP7/2007-2013) under grant agreement No. 335431 and by the DFG in the framework of SFB 925, GrK 1355 and the excellence cluster the Hamburg Centre for Ulrafast Imaging CUI.
L.W.C. is supported by the Grainger Graduate Fellowship. This work is supported by NSF MRSEC (Grant No. DMR-1420709), NSF Grant No. PHY-1511696, and ARO-MURI Grant No. W911NF-14-1-0003.

\section*{Acknowledgments}
We thank W. Weimer and K. Morgener for their contributions during the early stage of the experiment. 

\end{document}